\documentclass[preprint,showpacs,amsmath,amssymb,aps,prc,superscriptaddress]{revtex4}
\usepackage{epsfig}
\begin{document}

\preprint{APS/123-QED}

\title{ALPHA-CLUSTER STATES POPULATED IN $^{24}$MG + $^{12}$C}

\author{C. Beck$^1$, A. S\`{a}nchez i Zafra$^1$, P. Papka$^1$,
S. Thummerer$^1$, F. Azaiez$^1$, P. Bednarczyk$^1$, S. 
Courtin$^1$, D. Curien$^1$, O. Dorvaux$^1$, D. Lebhertz$^1$,
A. Nourreddine$^1$, J. Robin$^1$, M. Rousseau}


\affiliation{Institut Pluridisciplinaire Hubert Curien - D\'epartement de
Recherches Subatomiques, UMR7178, 
IN2P3-CNRS et Universit\'{e} Louis Pasteur (Strasbourg I), 23 rue du Loess, 
B.P. 28, F-67037 Strasbourg Cedex 2, France }

\author{W. von Oertzen$^{ 2}$, B. Gebauer$^{ 2}$, Tz. Kokalova$^{ 2}$, C. Wheldon}

\affiliation{Hahn-Meitner-Institut, Glienicker Str. 100, D-14109 
Berlin, Germany}

\author{G.~de~Angelis$^{3}$, A. Gadea$^{3}$, S. Lenzi$^{3}$, S. Szilner$^{3}$, 
D.~R Napoli}

\affiliation{NFN-Lab. Nationali di Legnaro and Dipartimento di 
Fisica, I-35020 Padova, Italy}
 
\author{W.~N. Catford}
 
\affiliation{School of Physics and Chemistry, University of Surrey, 
Guildford, Surrey, GU2 7XH, UK}

\author{D.~G. Jenkins}

\affiliation{Department of Physics, University of York, York, 
YO10 5DD, UK}

\author{G. Royer}

\affiliation{Subatech, IN2P3-CNRS et Universit\'e-Ecole des Mines, 
4 rue A. Kastler, F-44307 Nantes Cedex 3, France}

\date{\today} 

\begin{abstract}

{Charged particle and $\gamma$-decays in light $\alpha$-like nuclei are 
investigated for $^{24}$Mg+$^{12}$C. Various theoretical predictions 
for the occurence of superdeformed and hyperdeformed bands associated with 
resonance structures with low spin are presented. The inverse kinematics 
reaction $^{24}$Mg$+^{12}$C is studied at E$_{lab}$($^{24}$Mg) = 130 MeV. 
Exclusive data were collected with the Binary Reaction Spectrometer in coincidence 
with {\sc EUROBALL IV} installed at the {\sc VIVITRON} Tandem facility at 
Strasbourg. Specific structures with large deformation were selectively populated 
in binary reactions and their associated $\gamma$-decays studied. Coincident 
events from $\alpha$-transfer channels were selected by choosing the 
excitation energy or the entry point via the two-body Q-values. The analysis of 
the binary reaction channels is presented with a particular emphasis on 
$^{20}$Ne-$\gamma$ and $^{16}$O-$\gamma$ coincidences.}

\end{abstract}


\maketitle

\section{Introduction}

The observation of resonant structures in the excitation functions for various 
combinations of light $\alpha$-cluster (N = Z) nuclei in the energy regime from 
the barrier up to regions with excitation energies of E$_{X}$ = 20-50~MeV remains 
a subject of contemporary debate~\cite{1,2}. These resonances have been 
interpreted in terms of nuclear molecules~\cite{1}. The question whether quasimolecular 
resonances represent true cluster states in the compound systems, or whether 
they simply reflect scattering states in the ion-ion potential is still 
unresolved~\cite{1,2}. In many cases, these resonant structures have been 
associated with strongly-deformed shapes and with clustering phenomena, 
predicted from the cranked $\alpha$-cluster model~\cite{3}, 
Hartree-Fock calculations~\cite{4} and the Nilsson-Strutinsky 
approach~\cite{5}. Of particular interest is the relationship between 
superdeformation (SD) and nuclear molecules, since nuclear 
shapes with major-to-minor axis ratios of 2:1 have the typical ellipsoidal 
elongation for light nuclei~\cite{6}. Furthermore, the structure of possible 
octupole-unstable 3:1 nuclear shapes - hyperdeformation (HD) - for actinide
nuclei has also been widely discussed~\cite{6,7} in terms of clustering 
phenomena. 

Large (quadrupole) deformations and $\alpha$-clustering in light N = Z nuclei 
are known to be general phenomena at low excitation energy. For high angular 
momenta and higher excitation energies, very elongated shapes are expected to 
occur in $\alpha$-like nuclei for A$_{\small CN}$ = 20-60. These predictions 
come from the generalized liquid-drop model, taking into account the 
proximity energy and quasi-molecular shapes~\cite{8}. In fact, highly deformed 
shapes and SD rotational bands have been recently discovered in several such 
N = Z nuclei, in particular, $^{36}$Ar and $^{40}$Ca using $\gamma$-ray 
spectroscopy techniques~\cite{9,10}. HD bands in $^{36}$Ar 
and its related ternary clusterizations are predicted theoretically  
\cite{11}. With the exception of the cluster decay of $^{56}$Ni 
recently studied using charged particle spectroscopy~\cite{12}, no evidence 
for ternary breakup has yet been reported \cite{13} in light nuclei; the 
particle decay of $^{36}$Ar SD bands (and other highly excited bands) is still 
unexplored. The main binary reaction channels of the $^{24}$Mg+$^{12}$C 
reaction, for which both resonant effects and orbiting phenomena \cite{13} 
have been observed, is investigated in this work by using charged 
particle-$\gamma$-ray coincidence techniques. 

\newpage

\section{Experimental results.}

\begin{figure}[th]
\centerline{\psfig{figure=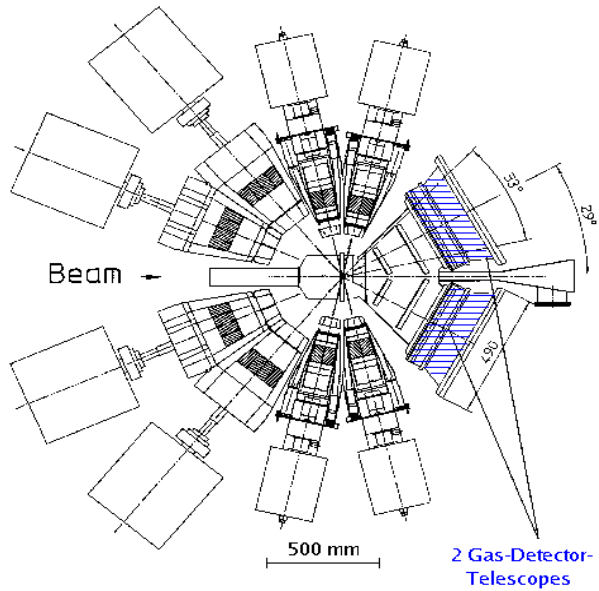,width=8cm,height=8cm}}
\vspace*{8pt}
\caption{\label{fig1} Schematic drawing of the scattering chamber 
showing the BRS arrangement for fragment detection and the $\gamma$-ray detectors 
of EB. At forward angles the two BRS gas detector telescopes are depicted, as well 
as two rings of Clover-Ge-detectors at angles around $\theta$ = 90$^\circ$ and 
Cluster-Ge-detectors at backward angles, respectively.}
\end{figure}

\begin{figure}[th]
\centerline{\psfig{figure=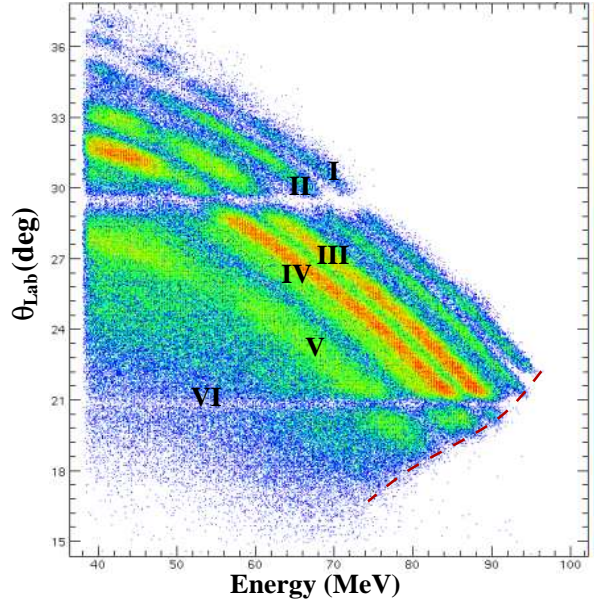,width=8cm,height=8cm}}
\vspace*{8pt}
\caption{\label{fig2} Two-dimensional angle versus energy spectrum, 
using fragment-fragment coincidences, measured for the $^{16}$O+$^{20}$Ne 
exit-channel. The relative intensity is shown on the side bar. The dashed line 
corresponds to the high-energy cutoff due to the geometrical bias of the kinematical 
coincidences. The regions labelled I to VI are defined in the text.}
\end{figure}

The study of charged particle-$\gamma$-ray coincidences in binary reactions in 
inverse kinematics is a unique tool in the search for extreme shapes related to 
clustering phenomena. In this paper, we investigate the $^{24}$Mg+$^{12}$C reaction 
with high selectivity at a bombarding energy E$_{lab}$($^{24}$Mg) = 130 MeV 
by using the Binary Reaction trigger Spectrometer (BRS) \cite{10,12} in 
coincidence with the {\sc EUROBALL IV} (EB) $\gamma$-ray spectrometer~\cite{10} 
installed at the {\sc VIVITRON} Tandem facility of Strasbourg. 
The $^{24}$Mg beam was produced and accelerated by the {\sc VIVITRON}, 
negative MgH$^{-}$ ions were extracted from the ion source and then the MgH 
molecules were cracked at the stripping foils of the terminal accelerator. The 
beam intensity was kept constant at approximately 5 pnA. The targets consisted 
of 200 $\mu$g/cm$^2$ thick foils of natural C. The choice of the 
$^{12}{\rm C}(^{24}{\rm Mg},^{12}{\rm C})^{24}{\rm Mg^{*}}$ reaction implies 
that for an incident beam energy of E$_{lab}$ = 130~MeV an excitation energy 
range up to E$^{*}$ = 30~MeV in $^{24}$Mg is covered.
The BRS, in conjunction with EB, gives access to a novel approach for the study of 
nuclei at large deformations as described below.

The BRS associated with EB combines as essential elements two large-area (with 
a solid angle of 187 msr each) heavy-ion gas-detector telescopes in a kinematical 
coincidence setup at forward angles. A schematic lay-out of the actual 
experimental set-up of the BRS with EB is shown in Fig.~1. The two telescope 
arms are mounted symmetrically on either side of the beam axis, each covering 
the forward scattering angle range 12.5$^\circ$-45.5$^\circ$, i.e. 
$\theta$ = 29$^\circ$ $\pm$ 16.5$^\circ$. For this reason the 30 tapered Ge 
detectors of EB~\cite{10} were removed.

Fig.~2 illustrates a typical example of a two-dimensional angle versus energy 
spectrum for the $^{16}$O+$^{20}$Ne exit-channel. The six regions labelled I to
VI have been defined as a function of the inelasticity of the reaction channel
from the ground-state Q-value E$^{*}$ = 0 (quasi-elastic) to full damping with 
E$^{*}$ larger than 12 MeV (orbiting deep-inelastic). The properties of
this $\alpha$-transfer channel will be further discussed thereafter.  

\newpage

\section{Discussion and conclusions}

\begin{figure}[th]
\centerline{\psfig{figure=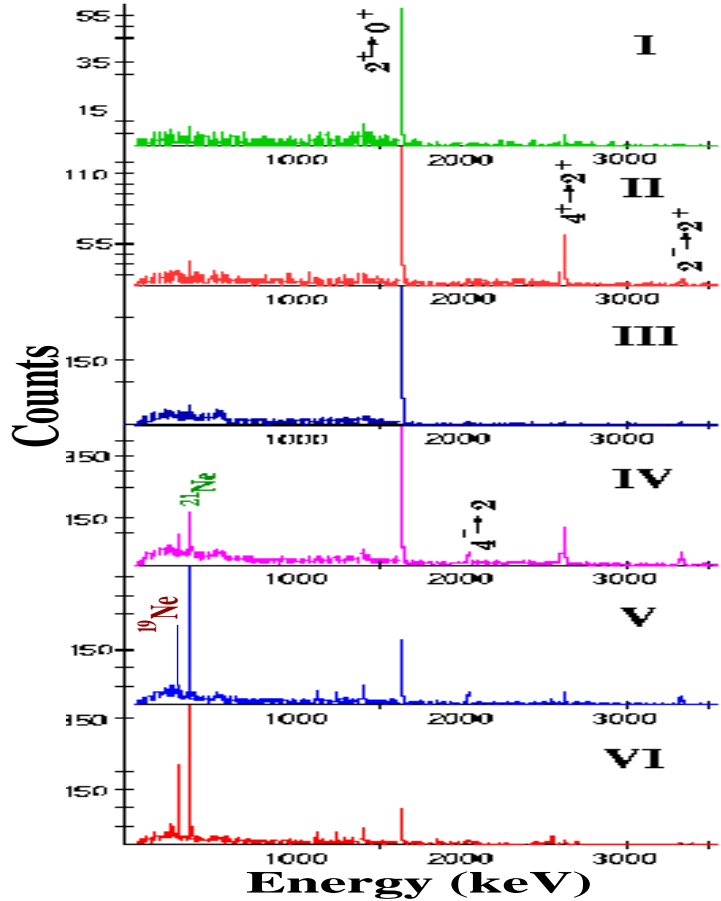,width=9.5cm,height=12cm}}
\vspace*{8pt}
\caption{\label{fig3} Gated $\gamma$-ray spectra, using 
fragment-fragment-$\gamma$-ray coincidences, measured for the $^{16}$O+$^{20}$Ne 
exit-channel. The six excitation energy gates labelled I to VI in 
Fig.~2 and defined in the text have been used as triggers to the six $\gamma$-ray 
spectra. The main $\gamma$-ray transitions in $^{20}$Ne are labelled.}
\end{figure}

Figs.~3 and 4 display the Doppler-shift corrected $\gamma$-ray spectra for events in 
coincidence with Z=8 and Z=10 gates defined in the BP vs E spectra as well as
gates on Fig. 2. 
Most of the known transitions of both $^{16}$O and $^{20}$Ne can be identified in the 
energy range depicted. The six different excitation energy gates displayed in Fig.~2 
are used to generate the $\gamma$-ray spectra shown in Fig.~3 (low-energy 
transitions). The $\gamma$-ray spectrum of Fig.~4 was 
triggered with the use of the gate labelled IV.

Identifications of the most intense $\gamma$ rays in $^{20}$Ne is straightforward 
and their labelling are given in Fig.~3. As expected, we observe decays feeding 
the yrast line of the $^{20}$Ne nucleus. Two previously unobserved transitions in 
$^{16}$O from the decay of the 3$^{+}$ state at 11.09~MeV are clearly visible in the 
$\gamma$-ray spectrum of Fig.~4, have been identified for the first time in 
Fig.~4 (inset) for new the partial level scheme. We note that, thanks to the excellent resolving power
of the EB+BRS set-up, the respective first escape peak positions of the 6.13 MeV, 
6.92 MeV and 7.12 MeV $\gamma$-ray transitions in $^{16}$O are also apparent in 
this spectrum. 

With appropriate Doppler-shift corrections applied to oxygen fragments identified 
in the BRS, it has been possible to extend the knowledge of the level scheme of 
$^{16}$O at high energies~\cite{14,15,16}, well above the 
$^{12}$C+$\alpha$ threshold, which is given in Fig.~4 for the sake of 
comparison. New information has been deduced on branching ratios of the decay of 
the 3$^{+}$ state of $^{16}$O at 11.085~MeV $\pm$ 3 keV (which does not 
$\alpha$-decay because of non-natural parity \cite{16}, in contrast to the 
two neighbouring 4$^{+}$ states at 10.36~MeV and 11.10~MeV, respectively) to the 
2$^{+}$ state at 6.92~MeV (54.6 $\pm$ 2 $\%$) and a value for the decay width 
$\Gamma_{\gamma}$ fifty times lower than the one given in the literature 
\cite{14,15}, it means $\Gamma_{3^+}$ $<$ 0.23 eV. This result is important 
as it is the highest known $\gamma$-decaying level for the well studied $^{16}$O nucleus 
\cite{14,15}. 

The connection of $\alpha$-clustering, quasimolecular resonances, orbiting phenomena 
and extreme deformations (SD, HD, ...) can be discussed in terms of 
the aspects of $\gamma$-ray spectroscopy of binary fragments 
from either inelastic excitations and direct transfers (with small energy damping
and spin transfer) or from orbiting (fully damped) processes \cite{13}.
Exclusive data were collected with the Binary Reaction Spetrometer (BRS) in 
coincidence with {\sc EUROBALL~IV} installed at the {\sc VIVITRON} Tandem facility 
of Strasbourg. New $\gamma$-ray spectroscopy results on $^{16}$O from the direct alpha transfer reactions has been
presented in this work.
The search for extremely elongated configurations (HD) in rapidly rotating 
medium-mass nuclei, which has been pursued exclusively using $\gamma$-spectroscopy, 
will have to be performed in conjunction with charged particle spectroscopy in the 
near future.

\newpage

\begin{figure}[th]
\centerline{\psfig{figure=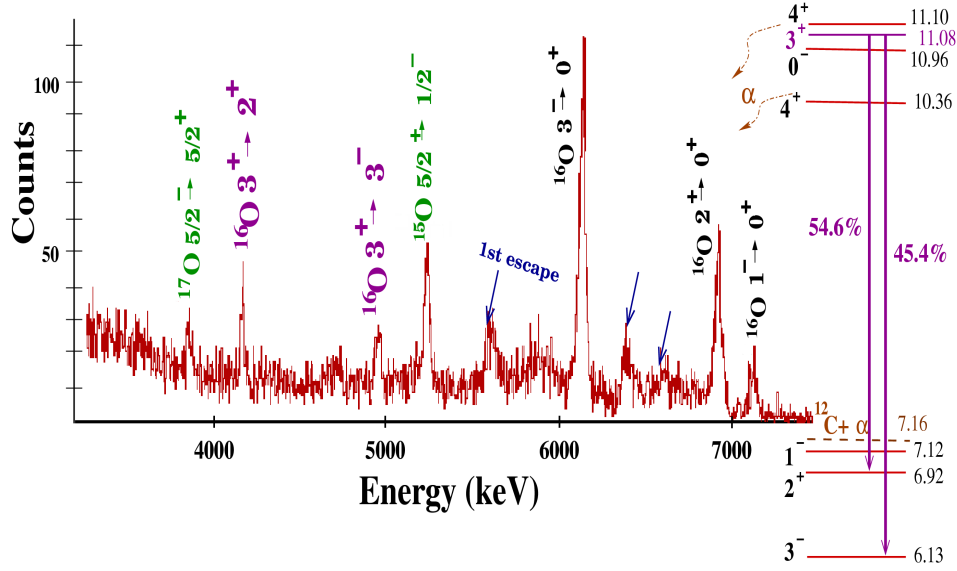,width=13cm,height=7.5cm}}
\vspace*{8pt}
\caption{\label{fig4}  $^{16}$O high-energy excited states populated 
in the $^{16}$O+$^{20}$Ne exit-channel with the gate IV of Fig.~2 as defined in the 
text. Doppler-shift corrections have been applied for O fragments detected in the 
BRS. The three blue arrows show the respective first escape peak positions of the 
6.13 MeV, 6.92 MeV and 7.12 MeV $\gamma$-ray transitions in $^{16}$O. The new 
partial level scheme of $^{16}$O is plotted in the inset}
\end{figure}

\medskip

\noindent
{\small
{\bf Acknowledgments:} We thank the staff of the {\sc VIVITRON} for providing us 
with good $^{24}$Mg stable beams and the EUROBALL group of Strasbourg for the 
excellent support in carrying out the experiment. This work was supported by the 
french IN2P3/CNRS, the german ministry of research (BMBF grant under contract 
Nr.06-OB-900), and the EC Euroviv contract HPRI-CT-1999-00078. S.T. would like to 
express his gratitude and warm hospitality during his three month stay in Strasbourg 
to the IReS and, he is grateful for the financial support obtained from the IN2P3, 
France. D.G.J. acknowledges receipt of an EPSRC Advanced fellowship.}

\newpage

\begin{center}

\end{center}

\end{document}